# Far-field radiation pattern in Coherent Anti-stokes Raman Scattering (CARS) Microscopy.

David Gachet, Nicolas Sandeau, Hervé Rigneault[*]
Institut Fresnel, Mosaic team, Domaine Univ. St Jérôme, 13397 Marseille Cedex 20, France

## ABSTRACT

Far field radiation pattern under tight focusing condition is investigated in Coherent Anti-stokes Raman Scattering (CARS) microscopy both in the forward (F-CARS) and backward (E-CARS) directions. While we assume no refraction index mismatch between the sample and the environing medium, our rigorous numerical electromagnetic computation takes into account the exact polarizations of the excitation laser beams and of the induced nonlinear dipoles. F-CARS and E-CARS radiation patterns, as well as their divergence, are studied as a function of the size of the sample object and compared to the excitation beams.

## 1. INTRODUCTION

Coherent Anti-Stokes Resonant Scattering (CARS) microscopy represents a unique approach to image chemical and biological samples using molecular vibrations as a contrast mechanism. CARS is a four-wave mixing nonlinear process[1] in which two laser fields of pulsations $\omega_p$ (pump) and $\omega_s$ (Stokes) interact with a medium to generate a field at the pulsation $\omega_{as}=2\omega_p-\omega_s$ (anti-Stokes). This field is dramatically enhanced when the pulsation difference $\omega_p-\omega_s$ equals the pulsation of a molecular vibration in the medium. First implemented in microscopy by Zumbusch[2] under tight focusing condition and collinearly propagating pump and Stokes beams, CARS microscopy appears nowadays as a powerful contrast mechanism to study living matter with chemical selectivity, reduced photodamage and three-dimensional sectioning capability[3].

Because of the coherent nature of the CARS contrast, the image formation process relies on multiple interferences between the elementary non linear dipoles that are photo-induced into the sample. This is very different from fluorescence microscopy where no interferences take place due to the incoherent nature of the fluorescence emission. The CARS coherent emission, that occurs where the excitation beams ($\omega_p$ and $\omega_s$) are focused and overlap, is also responsible for the asymmetric CARS emission in forward (F-CARS) and backward directions (E-CARS)[4].

Electromagnetic modelling of the CARS emission process under tight focusing condition have been reported[5] in the framework of a Green analysis and have focused mainly on the field in the direct space (x,y,z) and in the very vicinity of the object. In this paper we pursue an alternative way describing the CARS far field emission as an ensemble of coherently induced Hertzian dipoles located inside the object. This approach permits to compute the radiated anti-Stokes field ($\omega_{as}$) both in the direct (x,y,z) and in the reciprocal space ($k_x$, $k_y$, $k_z$). This model allows us to revisit the far field radiation patterns of beads with various diameters both in the forward and backward directions. Moreover, for the first time these patterns are presented in the **k** space. Eventually, the directivity of the CARS emission is compared with the incident beam's.

## 2. THEORY

We have implemented a full vector electromagnetic model[6] of both the focused incident fields and the emitted CARS radiation in a 3D object. A collinear geometry, where the pump and the Stokes beams overlap and focus through the same objective lens, is considered. This objective lens is assumed to be an achromatic thin lens.

### 2.1. Excitation field

---

[*] herve.rigneault@fresnel.fr; http://www.fresnel.fr/mosaic/

Both laser beams (pump and Stokes) are assumed to be Gaussian, monochromatic and linearly polarized along the x axis. The divergence of the laser beams can be neglected so that, in a plane before the objective (see fig. 1), the electric field $\mathbf{E_i}$ is given by

$$\mathbf{E_i} \propto \exp\left[-\left(\frac{h}{\sigma_g}\right)^2\right]\mathbf{e_x}$$

where $\sigma_g$ is the full-width half-maximum (FWHM) of the Gaussian beam. One can note on figure 1 that

$$h = f \sin\theta \qquad (2)$$

where f is the focal length of the objective. Introducing the ratio[7] $\beta$ of the objective back-aperture (r) to the beam radius ($\sigma_g$) proposed by Hess et al. ($\beta = r/\sigma_g$), $\mathbf{E_i}$ can be written

$$\mathbf{E_i}(\theta) \propto \exp\left[-\left(\frac{n\beta\sin\theta}{\text{NA}}\right)^2\right]\mathbf{e_x} \qquad (3)$$

where NA is the numerical aperture of the objective and n is the refractive index of the medium in the object space. The objective lens transforms the plane P in a reference sphere SO in the object space. Thus, introducing the geometrical optics intensity law proposed by Richards et al.[8], the electric field $\mathbf{E_s}$ on SO is given by

$$\mathbf{E_s} = \sqrt{n\cos\theta}\,\exp\left[-\left(\frac{n\beta\sin\theta}{\text{NA}}\right)^2\right]\mathbf{e_s} \qquad (4)$$

where

$$\mathbf{e_s} = \frac{1}{\sqrt{\cos^2\theta + (\sin\theta\cos\varphi)^2}}\begin{pmatrix}\cos\theta \\ 0 \\ -\sin\theta\cos\varphi\end{pmatrix}. \qquad (5)$$

We assume no refractive index mismatch between the sample and the environing medium in the object space. Therefore, at each point M in this space, the electric field $\mathbf{E}$ of the focused laser beam is given by

$$\mathbf{E} = \iint_\Omega \mathbf{E_s}\exp\left[i(\mathbf{k}\cdot\mathbf{F_oM})\right]d\Omega \qquad (6)$$

where $F_o$ is the objective focal point, $\Omega$ the solid angle delimited by the numerical aperture of the objective and $\mathbf{k}$ the wave vector defined by

$$\mathbf{k} = \frac{2\pi}{\lambda}\begin{pmatrix}\sin\theta\cos\varphi \\ \sin\theta\sin\varphi \\ \cos\theta\end{pmatrix}, \qquad (7)$$

where $\lambda$ is the wavelength of the laser beam (pump or Stokes).

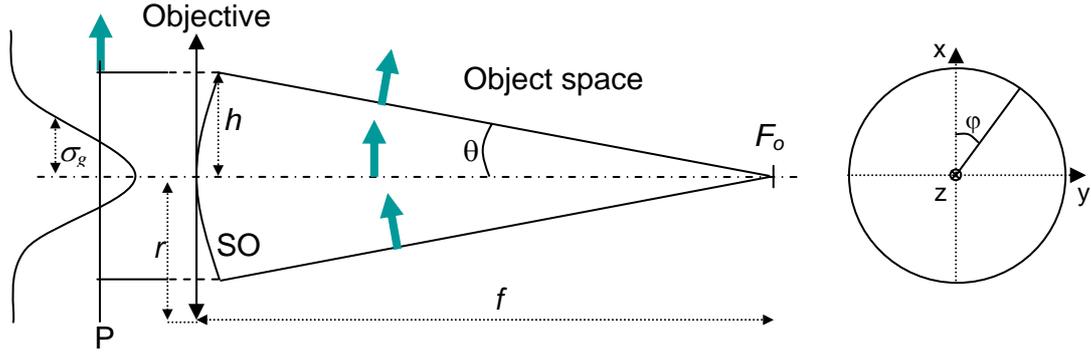

Fig. 1: Gaussian beam focused by an objective lens.

**2.2. Far field CARS emission**

The modeled object is assumed to be constituted of molecular species. Each chemical bond excited by the two laser beams can be considered to behave like a coherent dipole which polarization is given by

$$\mathbf{p} = \chi^{(3)} \mathbf{E}^2(\varpi_p) \mathbf{E}^*(\varpi_s) \qquad (8)$$

where **E** is the electric field introduced in equation (6) and

$$\chi^{(3)}_{ijkl} = \chi \left( \delta_{ij}\delta_{kl} + \delta_{ik}\delta_{jl} + \delta_{il}\delta_{jk} \right) \qquad (9)$$

$\chi$ is a complex number and $\delta$ is the Kronecker delta function. Thus equation (8) becomes

$$p_i = \chi \sum_{j=(x,y,z)} E_j(\varpi_p) \left( 2 E_i(\varpi_p) E_j^*(\varpi_s) + E_j(\varpi_p) E_i^*(\varpi_s) \right) \qquad \forall i = (x,y,z) \qquad (10)$$

Fig. 2 shows a Hertzian dipole possessing a polarization **p** and characterized by its far electric field $\mathbf{E_d}$ at a distant point M.

$$\mathbf{E_d} \approx \frac{-k_{as}^2}{4\pi \varepsilon_0 r} (\mathbf{p} \cdot \mathbf{e_r}) \exp(i k_{as} r) \mathbf{e}_\theta \qquad (11)$$

where r is the distance between the dipole and the point M, ($\mathbf{e_r}, \mathbf{e_\theta}, \mathbf{e_\phi}$) is the dipole spherical base (see Fig. 2) and $k_{as}$ is the wave number of the anti-Stokes radiation ($k_{as}=\omega_{as}/c$). Therefore, the illuminated sample object is considered as an assembly of coherent dipoles which polarization only depends on the position of the dipole (see Eq. 10). Thus, the total field emitted by the object can be calculated at each point of the object space. It is the sum of fields emitted by each dipole (see Eq. 11).

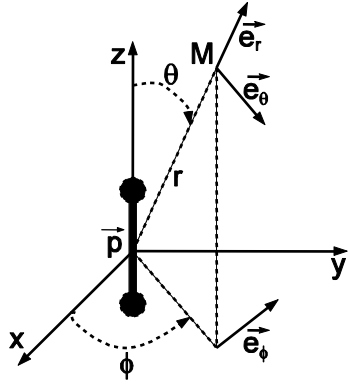

Fig. 2 : Hertzian dipole with a polarization **p**.

## 3. FAR FIED RADIATION PATTERN OF BEADS

We consider beads of various diameters embedded in a uniform medium of refractive index n=1.33 (water). The beads and the surrounding medium are assumed to have the same refractive index (no refractive effects) and to be non dispersive. The polarizations of the incoming beams (prior focusing) are considered linear along the x axis (Eq. 1). The incoming beams are focused in the bead center, the pump and Stokes beam have a wavelength of $\lambda_p$=750nm and $\lambda_s$=825nm respectively. The numerical aperture (NA) of the focusing lens is set to 1.2. With these values, the waists of the pump and Stokes beams along the x axis are 280nm and 300nm respectively.

### 3.1. Far field intensity in the direct space

Fig. 3 presents the far field intensity radiation patterns in the (x,z) plane of beads with various diameters d (100nm to 2μm), the color scale is not comparable between different diameters (see Fig. 4 for comparison). As already reported in Reference 4 the radiation pattern is dipole like for beads with a diameter much smaller than $\lambda_p$ (d=100nm) and the amount of radiated energy is similar in both forward (F-CARS) and backward (E-CARS) directions. This symmetry breaks down for larger bead diameters (d>200nm) and the F-CARS signal becomes rapidly much stronger than the E-CARS signal due to constructive interferences in the forward direction. As it is now well known, this effect as the practical consequence that CARS scatterers much smaller than $\lambda_p$ can be observed in the backward direction whereas CARS scatterers larger than $\lambda_p$ are easily observed in the forward direction.

Let us consider now the emission in the **k** space to appreciate the divergence of the emitted radiation.

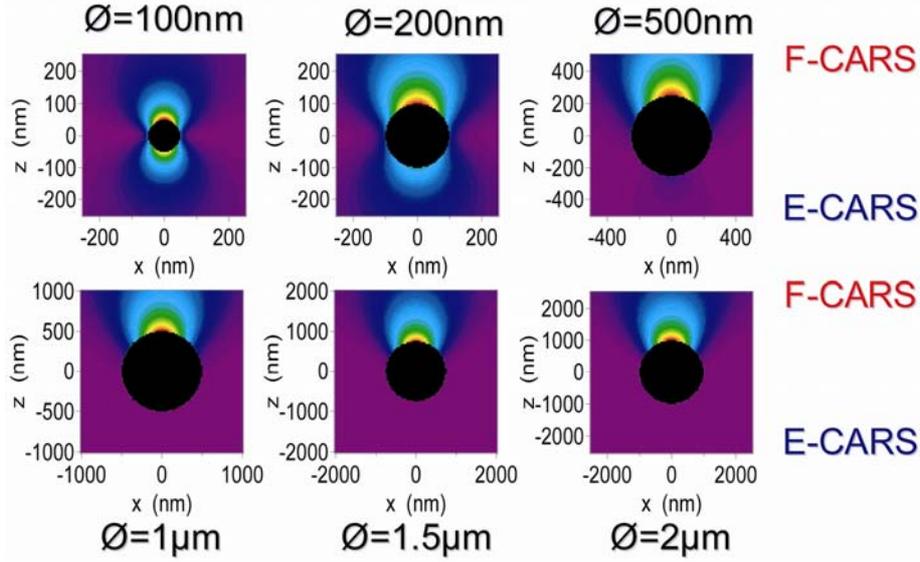

Fig. 3: Far field intensity radiation pattern in the (x,z) plane of beads with various diameters d (100nm to 2μm), the color scale is not comparable between different diameters. The pump and Stokes beams propagate along the z axis, they are focused at the bead center, polarized along the x axis (before focusing) and exhibit a waist along the x axis of 280nm and 300nm respectively.

### 3.2. Far field intensity in the reciprocal space

Fig. 4 gives the far field intensity radiation patterns in the $(k_x,k_y)$ space for beads of different diameters. The unit are in $k_x/k_0$ and $k_y/k_0$ where $k_0=2\pi/\lambda_{AS}$ and $\lambda_{AS}$ the anti-Stokes wavelength in vacuum. The color scales are different on each figure but can be compared thanks to the associated scales. For each diameter both the F-CARS and E-CARS radiation patterns are given; on the same figure, and for direct comparison, we have displayed the pump and Stokes beams in the **k** space (for this specific figure, the color scale is not comparable to the other CARS figures).

For a bead diameter d=100nm, one can recognize the intensity radiation pattern of a dipole lying along the x axis, the cone of emission being very broad along the y axis ($k_y/k_0$ reaches 1.2, a value much larger than the excitation beams). As already mentioned in Fig. 3., the emitted power in F-CARS and E-CARS are similar. When d≥500nm the E-CARS emission becomes negligible as compared to the F-CARS emission although subtle interference effects create some hot spots in the E-CARS **k** space. When d>500nm the radiation pattern extension along the $k_y$ axis reduces progressively to reach the value comparable to that of the excitation beams (case of the d=2μm bead). On the contrary, along the $k_x$ axis the divergence remains similar to the small diameter case.

This effect can be physically understood considering the CARS emission as a spatial transformation in the **k** space between the incident fields and a dipole emission. For a small bead diameter, one recovers the dipole emission with a 'narrow' divergence along the $k_x$ axis and a large one along the $k_y$ axis. For larger diameters, the similar phases of the induced dipoles along the y axis ruled out the dipole like radiation pattern along this axis and concentrate the radiation to recover ultimately the divergence of the excitation beams along this $k_y$ axis. Ultimately (bead of diameter d=2μm) this process generates a beam which has the divergence of the dipole emission along the $k_x$ axis (and therefore smaller than the excitation beams) and a divergence similar to the excitation beams along the $k_y$ axis. All in all, the CARS process is able to generate (for large diameter) a F-CARS beam which has a lower divergence than the excitation beams. From a practical point of view it is possible to use a low NA lens (see next section) to collect most of the F-CARS emitted radiation coming from objects bigger than few $\lambda_p$. Let us investigate more quantitatively this property in the next section.

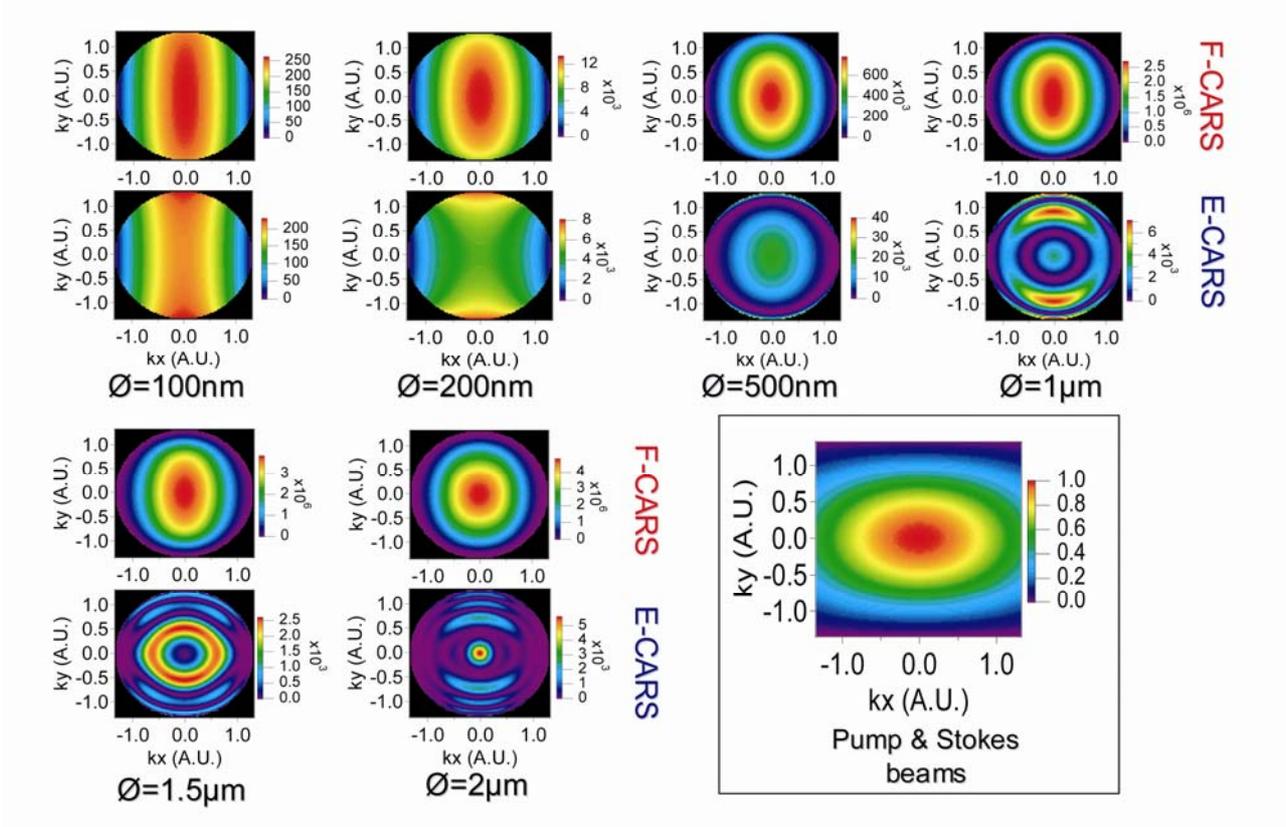

Fig. 4: F-CARS and E-CARS far field intensity radiation patterns in the ($k_x$,$k_y$) space for beads of different diameters. Unit in $k_x/k_0$ and $k_y/k_0$ ($k_0=2\pi/\lambda_{AS}$). The color scales are different on each figure but can be compared thanks to the associated scales. Inset: Pump and Stokes beams in the **k** space (scale not comparable to the other CARS figures).

### 3.3. F-CARS total radiated intensity

From the F-CARS intensity radiation patterns (Fig. 4.), it is possible to compute the amount of energy collected by a lens whose numerical aperture is $NA_F$. In the considered collinear geometry this lens is used only for the F-CARS signal collection. Fig. 5 plots the amount of F-CARS collected power for beads of different diameters with $NA_F$. We have normalized to 1 the total energy emitted for each bead in a $NA_F=1.2$. On the same graph, and for comparison, we have plotted the curve of the excitation pump and Stokes beams. The feature addressed in the previous section concerning the energy concentration in the **k** space for large bead diameter is clearly visible. For beads with diameter d≥500nm, the curve are all above the 'excitation beams' curve telling that a specific fraction of the total emitted F-CARS radiation can be collected with a lower $NA_F$ than for the 'excitation beams'. Ultimately the total F-CARS emitted radiation of the d=2μm bead can be collected with $NA_F=0.85$, a value much smaller than the NA=1.2 of the focusing objective lens. Practically it has been suggested[9] to use the condenser lens ($NA_F$ ~0.5) of a conventional inverted microscope to collect the F-CARS emitted signal. From our computations such a $NA_F$ leads to a 60% collection efficiency for objects bigger than $2.5\lambda_p$ (2μm), a value viable considering the strong F-CARS signal generated by such big objects.

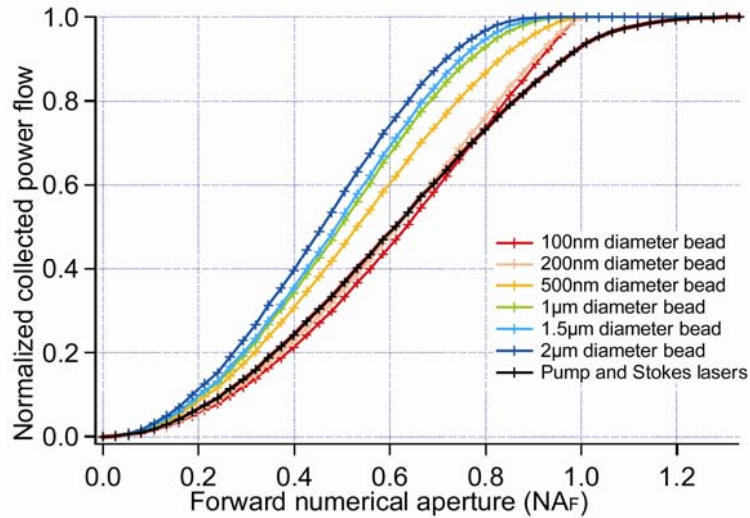

Fig.5: F-CARS collected power for beads of different diameters with $NA_F$. The total energy emitted for each bead in an $NA_F$=1.2 is normalized to 1. On the same graph, we have plotted the curve of the excitation pump and Stokes beams.

## 4. CONCLUSION

In this paper we have presented a rigorous electromagnetic modeling of CARS emission, considering both the excitation pump and Stokes fields and the CARS emission in a 3D object as vectorial. The presented model considers the objects as an ensemble of Hertzian dipoles photo-induced through the third order CARS non linear process. From such a description we have computed the far field CARS radiation pattern in both the forward and the backward directions for beads of different diameters. Our analysis in the **k** space has revealed how the CARS process can generate a F-CARS beam with a smaller divergence than the excitation beams. This surprising effect relies on the coherent constructive interference which takes place between the induced dipoles for bead diameter larger than few $\lambda_p$. In practice, such an effect permits to use low NA lens to collect most of the F-CARS emitted radiation.

## REFERENCES

[1] Y.R. Shen, *The principles of non linear optics*. (Wiley, New York, 1984).
[2] A. Zumbusch, G.R. Holtom, and X. S. Xie, "Vibrational microscopy using coherent anti-Stokes Raman scattering," Phys. Rev. Lett. **82**, 4014-4017 (1999).
[3] Andreas Volkmer, "Vibrational imaging and microspectroscopies based on coherent anti-Stokes Raman scattering microscopy," J. Phys. D: Appl. Phys. **38**, R59-R81 (2005).
[4] Andreas Volkmer, Ji-Xin Cheng, and X. Sunney Xie, "Vibrational Imaging with High Sensitivity via Epidetected Coherent Anti-Stokes Raman Scattering Microscopy," Phys. Rev. Lett. **87**, 23901 (2001).
[5] E.O. Potma, W.P. de Boeij, and D.A. Wiersma, "Nonlinear coherent four-wave mixing in optical microscopy," J. Opt. Soc. Am. B **17** (10), 1678-1684 (2000); Ji-Xin Cheng, Andreas Volkmer, and X. Sunney Xie, "Theoretical and experimental characterization of coherent anti-Stokes Raman scattering microscopy," JOSA B **19**, 1363-1375 (2002).
[6] N. Sandeau, *4Pi-microscopie: Applications à la localisation axiale de luminophore et à l'amélioration de la résolution latérale.* PhD thesis 2005AIX30031, Université Paul Cézanne Aix -Marseille III.
[7] S.T. Hess and W.W. Webb, "Focal volume optics and experimental artifacts in confocal fluorescence correlation spectroscopy", Biophysical Journal, **83** (4), 2300-2317 (2002).
[8] B. Richards and E. Wolf, "Electromagnetic diffraction in optical systems. II. Structure of the image field in an aplanatic system", Proceedings of the Royal Society of London A, **253**, 358-379 (1959).
[9] J. X. Cheng, Y. K. Jia, G. Zheng et al., "Laser-scanning coherent anti-Stokes Raman scattering microscopy and applications to cell biology," Biophys J **83** (1), 502-509 (2002)